\documentclass[a4paper,12pt]{article}
\usepackage{graphicx,amssymb,bm,latexsym,color,epsf}
\usepackage{cite}
\usepackage{amsmath}
\usepackage{appendix}
\usepackage[all]{xy}

\makeatletter
\@addtoreset{equation}{section}

\makeatother
\newcommand{\bea}{\begin{eqnarray}}
\newcommand{\eea}{\end{eqnarray}}
\newcommand{\vs}[1]{\vspace{#1 mm}}
\newcommand{\hs}[1]{\hspace{#1 mm}}
\renewcommand{\a}{\alpha}

\renewcommand{\d}{\delta}

\newcommand{\ve}{\varepsilon}
\newcommand{\s}{\sigma}

\newcommand{\la}{\lambda}
\newcommand{\pa}{\partial}
\newcommand{\nn}{\nonumber\\}
\newcommand{\p}[1]{(\ref{#1})}

\newcommand{\tlh}{\tilde h}

\newcommand{\VEV}[1]{\langle0| #1 |0\rangle}

\newcommand{\B}{{\rm B}}

\newcommand{\FP}{{\rm FP}}
\newcommand{\T}{{\rm T}}

\newcommand{\calB}{{\cal B}\kern-.70em{\cal B}}
\newcommand{\calC}{{\cal C}\kern-.58em{\cal C}}

\newcommand{\half}{(\kern-.3pt1\kern-.8pt/\kern-.2pt2)}

\newcommand{\mbf}[1]{{\boldsymbol #1}}
\newcommand{\rtg}{{\sqrt{-g}\,}}
\newdimen\Tdim
\def\Tspan#1{{\setbox0=\hbox{#1}%
\Tdim\ht0\advance\Tdim\dp0\advance\Tdim.55ex\rule[-\dp0]{0pt}{\Tdim}\box0}}

\begin{document}

\begin{flushright}
YITP-22-16
\end{flushright}
\medskip
\renewcommand{\thefootnote}{\fnsymbol{footnote}}

\begin{center}
{\large\bf
Covariant BRST Quantization of Unimodular Gravity II \\
\vs{3}
--- Formulation with a vector antighost ---
}
\vs{10}

{\large
Taichiro Kugo,$^{1,}$\footnote{e-mail address: kugo@yukawa.kyoto-u.ac.jp}
Ryuichi Nakayama,$^{2,}$\footnote{e-mail address: nakayama@particle.sci.hokudai.ac.jp}
and
Nobuyoshi Ohta\,$^{3,4,}$\footnote{e-mail address: ohtan@ncu.edu.tw}
} \\
\vs{5}

$^1$
{\em Yukawa Institute for Theoretical Physics, Kyoto University, Kyoto 606-8502, Japan}
\vs{3}

$^2$
{\em Division of Physics, Graduate School of Science,
Hokkaido University, Sapporo 060-0810, Japan}
\vs{3}

$^3$
{\em Department of Physics, National Central University, Zhongli, Taoyuan 320317, Taiwan}

and

$^4$
{\em Research Institute for Science and Technology,
Kindai University, Higashi-Osaka, Osaka 577-8502, Japan
}

\vs{5}
{\bf Abstract}
\end{center}

In our previous paper, we have presented a covariant BRST quantization of unimodular gravity
which may account for the smallness of the cosmological constant,
and we have shown that the physical degrees of freedom in the theory are the same as general relativity.
The formulation has been given by using rank-2 antisymmetric tensor fields for both ghosts and antighosts.
Here we give an alternative formulation using a vector field for the antighost
but keeping the same structure for the ghosts.
This gives a significantly simpler covariant quantization with less ghosts and no tripole modes in the ghost sector.
We show that this also gives only two physical transverse modes as in general relativity.

\renewcommand{\thefootnote}{\arabic{footnote}}
\setcounter{footnote}{0}

\newpage
\section{Introduction}

In our previous paper~\cite{KNO2}, hereafter referred to as I, we have presented a covariant local BRST quantization
of unimodular gravity (UG), and have shown that the physical degrees of freedom (dofs) in the theory are
two transverse modes as in general relativity (GR).
UG is an interesting theory that may explain why the cosmological constant is extremely
small~\cite{BD1,BD2,Unruh,HT,EVMU,NV}.

UG can be formulated as GR with the constraint that the determinant
of the metric should be a fixed volume form in the general relativity:
\bea
S_{\rm UG}= Z_N \int d^4 x \left[ \sqrt{-g} R + \la(\sqrt{-g}-\omega)\right],
\label{eq:UGaction2}
\eea
where $Z_N= 1/(16\pi G_N)$ with $G_N$ being the Newton constant, and $\lambda$ is a Lagrange multiplier
field to impose the constraint
\bea
\sqrt{-g}=\omega,
\label{unimodularity}
\eea
with $\omega$ being a fixed volume form.
Because of the unimodular constraint $\sqrt{-g}=\omega$, we can derive only the traceless part of
the Einstein equation even if there may be a ``cosmological constant'' in the action.
The real cosmological constant may be introduced as an integration constant, and thus is determined
by the boundary condition, not by a constant term in the action even if we have such a term.

The question how many physical dofs there exist in UG in the covariant quantization is a nontrivial problem and
there has been a lot of
debate~\cite{Smolin,FG,Eichhorn2013,Saltas,PS,Alvarez0,Alvarez1,BOT,Percacci2017,DOP,GM,HS,BP,DMPP,B,N,KNO1}.
The reason is the following. In the covariant BRST quantization of GR, there exist full diffeomorphism,
and we have four sets of ghosts and antighosts. This leaves $10-8=2$ dofs in GR.
However in UG, we have only transverse (or volume-preserving) diffeomorphism TDiff:
\begin{eqnarray}
\delta_{\B} g^{\mu\nu} & =& -\nabla^\mu c^\nu_{\T} -\nabla^\nu c^\mu_{\T}, 
\label{dbg} \\
\delta_{\B} \lambda&=& 0, 
\label{dbl}
\end{eqnarray}
expressed in terms of diffeomorphism Faddeev-Popov (FP) ghosts, $c^{\mu}_{\T}$,
which satisfies a transversality condition:
\begin{equation}
\nabla_\mu c^\mu_{\T}=0.  
\label{trans}
\end{equation}
The condition~\eqref{trans} eliminates one dof from the FP ghosts.
We would also have the same number of antighosts.
Consequently the BRST quantization of this system introduces only three sets of ghosts and antighosts.
We also have unimodular constraint, but it does not appear to introduce additional set of ghost and antighost.
Thus it seems that we are left with $10-6-1=3$ dofs, one more dof than GR.

In order to quantize UG in the covariant manner, we have to realize the symmetry~\eqref{dbg} off shell,
{\it i.e.} without using field equations. This is a nontrivial task, and it may appear to require
nonlocal projection operator~\cite{Alvarez1}.
However this is not the case, and it has long been known in supergravity\cite{deWit:1980lyi,Sohnius:1982fw,Kugo:1982cu}
that a vector subject to transverse constraint can be expressed by an unconstrained antisymmetric tensor even in
the curved spacetime.
In our previous paper I, based on this idea, we have expressed the reparametrization ghosts as
\begin{equation}
c^\mu_{\T} = \nabla_\nu c^{\nu\mu},
\label{eq:1by2}
\end{equation}
which automatically satisfy the transverse condition with rank-2 antisymmetric tensor ghost $c^{\mu\nu}$.
It turns out that after the first gauge fixing, the ghost system needs the ghosts for
ghosts~\cite{Townsend:1979hd,Kimura:1980zd,HKO}.
The reason is clear: The unconstrained rank-2 tensor has 6 dofs which are more than required
to express the transverse vector ghost modes with 3 dofs. We have found that this redundancy manifests itself
in the form of the gauge invariance in the ghost system, and this further requires the gauge fixing and
the introduction of ghosts, reducing the number of dofs.
Thus the ghost system becomes significantly more complicated than usual.
Since we must have the same number of the antighosts, it is natural to introduce similar rank-2 tensor antighosts,
and this further requires the ghosts for ghosts. We have to continue the gauge fixing and the introduction of
the ghosts until there remains no more gauge invariance.
The important discovery in I is that after all this gauge fixing of TDiff, the multiplier field $\la$ is actually
identified with a BRST daughter. This means that there exists a set of ghost and antighost corresponding to
the unimodular constraint after gauge fixing only TDiff, contrary to the above naive expectation.
This is the key observation to get the correct number of dofs.

This formulation is nice in the sense that it gives a formulation symmetric in ghosts and antighosts,
but use of the rank-2 antisymmetric tensors for both ghosts and antighosts gives complicated structure because of
the necessity of the ghosts for ghosts in both sectors.
Here we note that what is really required for the off-shell gauge fixing
of TDiff is to use the antisymmetric tensor fields only for the ghosts (not antighosts) to express the transverse
transformation parameter as in \eqref{eq:1by2}.
We further notice that, in our other paper~\cite{KNO1} for the quantization of GR in unimodular gauge,
we have actually presented a general way of imposing such a transverse-vector gauge condition
by using a ($d$-component) vector antighost. It is realized at the price of adding an extra scalar
field BRST doublet, a set of BRST parent and daughter.
The variation of the action by this BRST daughter field impose
the transverse condition on shell on the vector antighost, thus leaving the necessary number of dofs for antighosts.
Because the structure in the antighost sector in this formulation does not need the ghosts for ghosts, we expect
that this formulation gives considerably simpler covariant quantization of UG with less ghosts.
This is what we aim in this paper, and indeed we show that this formulation successfully gives an alternative
and simpler covariant BRST quantization with correct number of dofs. It turns out that this formulation
also has the advantage that  in the ghost and antighost sectors, there do not exist tripole modes, which
existed in I due to the use of the ghosts and antighosts with derivatives~\eqref{eq:1by2}.
The formulation, however, gives the asymmetric one in the ghost and antighost sectors.

This paper is organized as follows.
In sect.~\ref{brstq}, we start with the off-shell gauge fixing of TDiff using rank-2 antisymmetric tensor ghosts
and vector antighosts. The structure in the ghost sector is basically the same as in I, and we have the ghosts for
ghosts phenomena. We gauge fix TDiff until there remains no more invariance.
The structure for the antighost sector is similar to that in \cite{KNO1}.
In sect.~\ref{linear}, to study the spectrum in the theory, we concentrate on the theory at the linearized level.
First, to check that we have fully gauge fixed the gauge invariance, in subsect.~\ref{brstq.2}, we show that
there indeed exist the propagators for the flat background for simplicity.
Next, in subsect.~\ref{eom}, we derive the equations of motion (EOMs) at the linearized order.
We can see that there is simplification in the antighost sector, and there is no tripole field
in ghosts and antighosts in contrast to our pervious formulation in I
though there is tripole in the graviton excitation.
In sect.~\ref{counting}, we identify which fields represent independent modes.
We use this result in sect.~\ref{ident} to examine how most of the fields fall into the BRST quartets
and show that there remain only 2 physical dofs in the theory.
In sect.~\ref{discussions}, we summarize our results and conclude the paper with some discussions.

\section{BRST quantization of unimodular gravity }
\label{brstq}

The action (\ref{eq:UGaction2}) is invariant under the BRST transformation~\eqref{dbg} and \eqref{dbl}
expressed in terms of diffeomorphism FP ghosts, $c^{\mu}_{\T}$,
which satisfies a transversality condition~\eqref{trans}.

In our previous paper~I, we have expressed this field in terms of an unconstrained antisymmetric
rank-2 tensor ghost $c^{\mu\nu}$ as ~\eqref{eq:1by2}.
It was shown in detail that imposing the nilpotency of the BRST transformation automatically
clarifies the existence of additional gauge invariance and ghosts for ghosts~\cite{Townsend:1979hd,Kimura:1980zd,HKO}.
Here we just summarize the result, referring to I for the details.
The BRST transformation laws are
\begin{align}
\delta_{\B} c^{\nu\mu} &= c^\nu_{\T}c^\mu_{\T} + i\nabla_\rho d^{\rho\nu\mu} ,\nn
\delta_{\B} d^{\rho\nu\mu} &= i c^\rho_{\T}c^\nu_{\T}c^\mu_{\T} - \nabla_\sigma t^{\sigma\rho\nu\mu},
\label{ghostBRST}
\\
\delta_{\B} t^{\sigma\rho\nu\mu} &= ic^\sigma_{\T} c^\rho_{\T}c^\nu_{\T}c^\mu_{\T} . \nonumber
\end{align}
The field $d^{\rho\nu\mu}$ is a hermitian boson carrying double ghost number $N_\FP=+2$
and denotes the ghost for the ghost corresponding to the gauge transformation of $c^{\mu\nu}$
under which the ``field strength'' $c^\mu_{\T} = \nabla_\nu c^{\nu\mu}$ is invariant.
Another field $t^{\sigma\rho\nu\mu}$ is similarly introduced.

We now consider the BRST quantization of this UG system.
Here the crucial difference from our previous paper I is that we do not use the antisymmetric tensor field
as the multiplier BRST doublet field for fixing the $(d-1)({=}3)$-component TDiff gauge invariance.
Instead we use the method developed in \cite{KNO1}
to use a $d\,({=}4)$-component (unconstrained) vector multiplier doublet
field $(\bar{c}_\mu,\,b_\mu)$:
\begin{equation}
\delta_\B \bar{c}_\mu = i b_\mu\,,
\end{equation}
for fixing the $(d-1)$-component TDiff gauge at the price of adding an extra
scalar BRST doublet field $(S,\,C_S)$, transforming as
\bea
\delta_B S &=& C_S\,.
\eea

The gauge-fixing and FP ghost (GF+FP) terms in the first step are given by~\cite{KU,Ohta2020}
\begin{align}
{\cal L}_{\rm GF+FP,1} &= -i\d_{\B} \bigl[ \bar c_\mu(\pa_{\la} \tilde{g}^{\la\mu}-\tilde{g}^{\mu\nu}\pa_\nu S)\bigr]
\nn
&= b_\mu( \partial_{\lambda}\tilde{g}^{\lambda\mu}-\tilde g^{\mu\nu} \pa_\nu S)
+i\bar{c}_\mu\left[ \pa_{\nu} \d_{\B}(\tilde g^{\mu\nu})-\d_{\B}(\tilde g^{\mu\nu})\pa_\nu S
-\rtg \pa^\mu C_S \right],
\label{gfaction1}
\end{align}
where $\tilde g^{\mu\nu} \equiv\rtg g^{\mu\nu}$, and
\begin{align}
\delta_\B \tilde{g}^{\mu\nu}
&= -\rtg(\nabla^\mu c_\T^\nu+\nabla^\nu c_\T^\mu) + \tilde{g}^{\mu\nu}\nabla_\lambda c^\lambda_\T \nn
&= -\rtg(\nabla^\mu\nabla_\rho c^{\rho\nu}+\nabla^\nu\nabla_\rho c^{\rho\mu}) .
\end{align}
Here in Eq.~(\ref{gfaction1}),
we can see the double roles of the extra scalar $S$. First,
the gauge-fixing condition resulting from the variation of the multiplier field $b_\mu$ is
\begin{equation}
\partial_{\lambda}\tilde{g}^{\lambda\mu}-\tilde g^{\mu\nu} \pa_\nu S = 0,
\end{equation}
which demands that the 4-component de Donder gauge condition
$\partial_\lambda\tilde{g}^{\lambda\mu}=0$ be satisfied aside from
the ``longitudinal'' component $\tilde{g}^{\mu\nu}\partial_\nu S\equiv \rtg\partial^\mu S$
which remains arbitrary since $S$ is
nowhere else specified. So the presence of $\partial^\mu S$ term reduces the actual
number of gauge conditions on the metric field from 4 to 3.
Second, the variation of the BRST partner $C_S$ of $S$
gives the transverse constraint on the partner multiplier
$\bar{c}^\mu:=g^{\mu\nu}\bar{c}_\nu$ of $b^\mu:=g^{\mu\nu}b_\nu$:
\begin{equation}
\partial_\mu(\rtg g^{\mu\nu}\bar{c}_\nu) = \rtg \nabla_\mu\bar{c}^\mu= 0.
\end{equation}
The variation of $S$ field itself yields the BRST transform of this equation as the EOM:
\begin{equation}
\delta_\B[\partial_\mu(\rtg g^{\mu\nu}\bar{c}_\nu)]
= \rtg \nabla_\mu\delta_\B \bigl(g^{\mu\nu}\bar{c}_\nu\bigr)
= \rtg \bigl(\nabla_\mu b^\mu+\nabla_\mu\delta_\B(g^{\mu\nu})\bar{c}_\nu\bigr) = 0.
\end{equation}
Although $\delta_\B\bar{c}_\nu=b_\nu$, the contravariant multiplier
$b^\mu=g^{\mu\nu}b_\nu$ is required to be transverse on shell
up to FP ghost quadratic term $\nabla_\mu[\delta_\B(\tilde{g}^{\mu\nu})\bar{c}_\nu]$.

This ghost Lagrangian (\ref{gfaction1}) depends on $c^{\nu\mu}$ only
through $c^\mu_\T = \nabla_\nu c^{\nu\mu}$ and
has the gauge invariance under the transformations
with rank-3 totally antisymmetric parameters $\varepsilon^{\rho\nu\mu}$:
\begin{align}
\delta c^{\nu\mu} &= \nabla_\rho\varepsilon^{\rho\nu\mu}.
\label{eq:GT1-1}
\end{align}
This is just the gauge invariance already lifted in our BRST transformation
(\ref{ghostBRST}) with the ghost for ghost field $d^{\rho\nu\mu}$.
We take the the following gauge-fixing condition and introduce
a multiplier BRST doublet to impose it:
\begin{eqnarray}
\begin{array}{ccc}
\text{gauge fixing cond.}  &   :      &  \text{multiplier BRST doublet} \\
\nabla^{[\rho}c^{\nu\mu]} = 0  &  :& (\bar{d}_{\rho\nu\mu},\bar c_{\rho\nu\mu}  ),  \quad
\delta_{\B} \bar{d}^{\rho\nu\mu} = \bar c^{\rho\nu\mu}.
\end{array}
\label{eq:2ndGF}
\end{eqnarray}
Here and in what follows, the bracket $[\ ]$ attached to the indices means
the weight 1 antisymmetrization;
e.g., $A_{[\mu}B_{\nu]}=(1/2)(A_\mu B_\nu-A_\nu B_\mu)$.
Similarly we will also use $( \ )$ for the symmetrization with weight 1.

The GF+ FP terms in the second step are
\bea
{\cal L}_{\rm GF+FP,2}
&=& \frac{i}{2} \rtg  \delta_{\B} \big(\bar{d}_{\rho\nu\mu}\nabla^\rho c^{\nu\mu} \big)
{}= \frac{i}{2} \rtg  \delta_\B \big(\bar{d}^{\rho\nu\mu}\nabla_\rho c_{\nu\mu} \big)
\nn
&=&\frac{i}{2} \rtg \big[ \bar{c}^{\rho\nu\mu} \, \nabla_\rho c_{\nu\mu}
 - \nabla_{\rho}\bar{d}^{\rho\nu\mu} \cdot \delta_{\B} (g_{\nu\sigma}g_{\mu\kappa})c^{\sigma\kappa}
-\nabla^{\rho}\bar{d}_{\rho\sigma\kappa} \cdot  (c^{\sigma}_{\T}c^{\kappa}_{\T}+i\nabla_{\mu}d^{\mu\sigma\kappa})
 \big] ,\nn
\label{gfaction2}
\eea
where partial integrations have been performed in the second and third terms, and
use has been made of the commutativity
$\delta_\B\bigl(\nabla_{[\mu_1}A_{\mu_2\cdots \mu_n]}\bigr)
=\nabla_{[\mu_1}\delta_\B (A_{\mu_2\cdots \mu_n]})$ following from the equality
$\nabla_{[\mu_1}A_{\mu_2\cdots \mu_n]}= \partial_{[\mu_1}A_{\mu_2\cdots \mu_n]}$ valid for
any totally antisymmetric tensor $A_{\mu_2\cdots \mu_n}$.

This action (\ref{gfaction2}) still has the gauge invariance under the transformations~\cite{HKO}
\begin{align}
\delta d^{\rho\nu\mu}&=  \nabla_\sigma\varepsilon^{\sigma\rho\nu\mu} ,    \label{eq:GT2-1} \\
\delta\bar{d}^{\rho\nu\mu}&= \nabla_\sigma\bar\varepsilon^{\sigma\rho\nu\mu} , \label{eq:GT2-2} \\
\delta\bar{c}^{\rho\nu\mu}&= \nabla_\sigma\bar\theta^{\sigma\rho\nu\mu}  , \label{eq:GT2-4}
\end{align}
since it depends on these tensor fields only through their
covariant divergences like $\nabla_\rho d^{\rho\nu\mu}$,
if partial integration is performed when necessary. Here again, the first
gauge invariance (\ref{eq:GT2-1}) is the one already lifted in our BRST
transformation (\ref{ghostBRST}) with the ghost for ghost field
$-t^{\sigma\rho\nu\mu}$. The second gauge transformation (\ref{eq:GT2-2})
for the BRST parent field $\bar{d}^{\rho\nu\mu}$ is contained as
a part of the multiplier BRST transformation in Eq.~(\ref{eq:2ndGF}).
We fix the former two gauge invariances by the following
gauge-fixing conditions and introduce the corresponding multiplier BRST
doublets to impose them:
\begin{eqnarray}
\begin{array}{ccc}
\text{gauge fixing cond.}  &   :   &  \text{multiplier BRS doublet} \\
\nabla^{[\sigma}d^{\rho\nu\mu]} = 0  &:& (\bar{t}_{\sigma\rho\nu\mu},\bar{d}_{\sigma\rho\nu\mu}), \quad
\delta_{\B} \bar{t}^{\sigma\rho\nu\mu} = i\bar{d}^{\sigma\rho\nu\mu}, \\
\nabla^{[\sigma}\bar{d}^{\rho\nu\mu]} = 0 &:& (c^{\sigma\rho\nu\mu},d^{\sigma\rho\nu\mu} ), \quad
\delta_{\B} c^{\sigma\rho\nu\mu} = i d^{\sigma\rho\nu\mu}.
\end{array}
\label{eq:3rdGF}
\end{eqnarray}
The third gauge-invariance is automatically fixed by the gauge-fixing on
$\bar{d}_{\rho\nu\mu}$ of the
second gauge-invariance because of $\delta_\B \bar{d}^{\rho\nu\mu}=\bar{c}^{\rho\nu\mu}$.

The GF + FP Lagrangian in the third step is
\begin{align}
{\cal L}_{\rm GF+FP,3}
&= -i \delta_{\B} \Bigl[ \rtg \big[ -\frac{1}{6}
\bar{t}_{\sigma\rho\nu\mu}\bigl(\nabla^{\sigma}d^{\rho\nu\mu} + \frac{\alpha}{4} d^{\sigma\rho\nu\mu}\bigr)
+\frac{1}{6}c^{\sigma\rho\nu\mu}\nabla_{\sigma}\bar{d}_{\rho\nu\mu}
\big]\Bigr] \nn
&=  \frac{1}{6} \rtg \Big[
- \bar{d}^{\sigma\rho\nu\mu}\nabla_{\sigma}d_{\rho\nu\mu}
-\frac{\alpha}{4}\bar{d}^{\sigma\rho\nu\mu}d_{\sigma\rho\nu\mu}
+i\nabla^{\sigma}\bar{t}_{\sigma\rho\nu\mu} \cdot \bigl(-\nabla_\lambda t^{\lambda\rho\nu\mu}
+ic^\rho_{\T}c^\nu_{\T}c^\mu_{\T}\bigr) \nn
&\hspace{4em}{}
+d^{\sigma\rho\nu\mu}\nabla_{\sigma}\bar{d}_{\rho\nu\mu}
+ic^{\sigma\rho\nu\mu}\nabla_{\sigma}\bar{c}_{\rho\nu\mu}
+i \nabla_{\sigma} \bar{t}^{\sigma\rho\nu\mu} \cdot d^{\kappa\tau\la}
 \cdot \delta_{\B} (g_{\rho\kappa}g_{\nu\tau}g_{\mu\la}) \nn
& \hspace{3em}{}
-i\nabla_{\sigma}c^{\sigma\rho\nu\mu} \cdot \bar{d}^{\kappa\tau\la}\delta_{\B}(g_{\rho\kappa}g_{\nu\tau}g_{\mu\la})
\Big]   .
\label{gfaction3}
\end{align}
We have introduced a gauge parameter $\alpha$ for later convenience.
The gauge condition for the third gauge symmetry (\ref{eq:GT2-4}) follows
from the terms containing $c^{\sigma\rho\nu\mu}$ in (\ref{gfaction3}).
We see that our antighost system is drastically simplified compared with that in I.


Now there remains no further invariance and we expect that the system is
now fully gauge fixed. To avoid too many tensor suffices, however,
we rewrite the antisymmetric
tensor fields by their (Hodge) dual fields.
Our sequence of ghost fields, $c^{\mu\nu}, \ d^{\mu\nu\rho}$ and $t^{\mu\nu\rho\sigma}$
are expressed by their dual fields $C_{\mu\nu},\ D_\mu,\ T$ (generally
denoted by the corresponding uppercase letters) as
\begin{align}
\rtg c^{\mu\nu} &= \frac{1}{2}\varepsilon^{\mu\nu\rho\sigma} C_{\rho\sigma}  ,  \nn
\rtg d^{\mu\nu\rho}&=  \varepsilon^{\mu\nu\rho\sigma} D_\sigma ,   \nn
\rtg t^{\mu\nu\rho\sigma}&= \varepsilon^{\mu\nu\rho\sigma} T .
\end{align}
Our convention for the $\ve$ is $\ve^{0123}=+1$ and $\ve_{0123}=-1$.

The $3$ multiplier BRST doublets are expressed by their duals as
\begin{align}
& \rtg
\begin{pmatrix}
\bar{d}^{\mu\nu\rho} \\
\bar{c}^{\mu\nu\rho}
\end{pmatrix}
= - \varepsilon^{\mu\nu\rho\sigma}
\begin{pmatrix}
\bar{D}_\sigma\\
\bar{C}_\sigma
\end{pmatrix} , \quad \rtg
\begin{pmatrix}
\bar{t}^{\mu\nu\rho\sigma} \\
\bar{d}^{\mu\nu\rho\sigma}
\end{pmatrix}
= - \varepsilon^{\mu\nu\rho\sigma}
\begin{pmatrix}
\bar{T} \\
\bar{D}
\end{pmatrix} ,  %
\nn
&
\rtg
\begin{pmatrix}
c^{\mu\nu\rho\sigma} \\
d^{\mu\nu\rho\sigma}
\end{pmatrix}
= -\varepsilon^{\mu\nu\rho\sigma}
\begin{pmatrix}
C \\
D
\end{pmatrix} ,
\end{align}
Furthermore $c^\mu_{\T}$ should be understood  to represent
\begin{align}
c^\mu_{\T}& = \nabla_\nu c^{\nu\mu} = -\frac{1}{2\rtg} \varepsilon^{\mu\nu\rho\sigma} \partial_\nu C_{\rho\sigma}.
\end{align}

In terms of these dual fields,
the BRST transformations for the ghost fields are rewritten as follows:
\begin{align}
\delta_{\B} C_{\mu\nu} &= -\frac12 \rtg \varepsilon_{\mu\nu\rho\sigma}c^\rho_{\T}c^\sigma_{\T}
 + i(\partial_\mu D_\nu-\partial_\nu D_\mu), \nn
\delta_{\B} D_\mu&= \frac{i}{3!}\rtg \varepsilon_{\mu\nu\rho\sigma}c^\nu_{\T}c^\rho_{\T}c^\sigma_{\T}
 + \partial_\mu T, \nn
\delta_{\B} T &=- \frac{i}{4!}\rtg \varepsilon_{\mu\nu\rho\sigma}c^\mu_{\T}c^\nu_{\T}c^\rho_{\T}c^\sigma_{\T}.
\end{align}
The BRST transformations of multiplier BRST doublets are trivial for covariant vectors and scalars:
\begin{align}
& \delta_{\B} \bar{D}_\mu=\bar{C}_\mu,\qquad \delta_{\B} \bar{T}=i\bar{D}, \qquad \delta_{\B} C=iD.
\end{align}

By using these dual fields, the GF+FP ghost Lagrangians (\ref{gfaction1}), (\ref{gfaction2}) and
(\ref{gfaction3}) are rewritten  as
\begin{align}
{\cal L}_{\rm GF+FP,1}
&= b_\mu(\pa_{\la}\tilde{g}^{\la\mu}-\tilde{g}^{\mu\nu}\pa_\nu S) \nn
& \hspace{2em}  + i \bar c_\mu\Big[
\pa_\nu\nabla^{(\mu} \big( \ve^{\nu)\rho\s\la}
 \pa_\rho C_{\s\la}\big)
- \d_{\B}(\tilde{g}^{\mu\nu}) \pa_\nu S
- \tilde{g}^{\mu\nu} \pa_\nu C_S
\Big] ,
\label{gfaction1P}
\end{align}
\begin{align}
\hs{-25}
{\cal L}_{\rm GF+FP,2}
&=
i \delta_{\B}\Bigl[\rtg \bar{D}_\sigma\nabla_\rho C^{\rho\sigma}\Bigr]
 \nn
&=
\rtg \Big[ i\bar{C}^{\sigma}\nabla^\rho C_{\rho\sigma}
+\frac{3i}{4\rtg}\varepsilon^{\kappa\mu\nu\la} \nabla^{\rho} \bar{D}^{\sigma}\cdot
 \nabla_{[\rho}C_{\kappa\sigma]}\cdot \nabla_{\mu}C_{\nu\la} \nn
& \hspace{1em}{}
+ \big(\nabla^{\mu}\bar{D}^{\nu}-\nabla^{\nu}\bar{D}^{\mu}\big) \big(\nabla_{\mu}D_{\nu}
-ig^{\rho\la}C_{\rho\nu} \d_{\B} g_{\mu\la}\big) \Big]
 ,  \label{gfaction2P}
\end{align}
\begin{align}
\hs{-25}
{\cal L}_{\rm GF+FP,3}
&=
-i \delta_{\B}\Bigl[ \rtg \bigl(
\bar{T}\nabla_\sigma D^\sigma + \alpha\bar{T}D + C\nabla_\sigma\bar{D}^\sigma \bigr)\Bigr]
\nn
&=
\sqrt{-g} \Big[\bar{D}\nabla^{\mu}D_{\mu}+\alpha\bar{D}D +D \nabla_{\mu} \bar{D}^{\mu}
+iC\nabla_{\mu}\bar{C}^{\mu} \nn
 &\hspace{3em}{}
-\frac{1}{4}\nabla^{\sigma}\bar{T} \cdot \nabla_{[\sigma}C_{\nu\mu]}
\big( \nabla^{\mu}C^{\la\rho}\cdot \nabla^{\nu}C_{\la\rho}-4\nabla_{\la}{C^{\mu}}_{\rho} \cdot \nabla^{\nu}C^{\la\rho}
\nn
&\hspace{3em}{} +2\nabla_{\la}{C^{\mu}}_{\rho}\cdot \nabla^{\la}C^{\nu\rho}
-2\nabla_{\la} C^{\mu\rho} \cdot\nabla_\rho C^{\nu\la}\big)
-i\nabla^{\mu}\bar{T} \cdot \nabla_{\mu}T
 \nn
& \hspace{3em}{} +(i\nabla^{\mu}\bar{T} \cdot D^{\nu} +i\nabla^{\mu}C\cdot \bar{D}^{\nu}) \delta_{\B} g_{\mu\nu}
\Big]
,
\label{gfaction3P}
\end{align}
where
\begin{align}
\delta_{\B} g_{\mu\nu} &= -(\rtg)^{-1}g_{\la(\nu}\varepsilon^{\la\rho\sigma\tau}\nabla_{\mu)}
\nabla_{\rho}C_{\sigma\tau}\ .
\end{align}
is to be substituted in the above equations.

\section{Propagators and equations of motion at linear order}
\label{linear}

\subsection{Propagators}
\label{brstq.2}

Now the total Lagrangian of our UG system is given by
\begin{equation}
{\cal L}_{\rm UG} = \rtg R + \lambda(\rtg - \omega) +
{\cal L}_{\rm GF+FP,1} (\ref{gfaction1P}) +
{\cal L}_{\rm GF+FP,2} (\ref{gfaction2P}) +
{\cal L}_{\rm GF+FP,3} (\ref{gfaction3P})\,.
\label{eq:totalL}
\end{equation}
Let us check in detail if we get nonsingular fully gauge fixed action on the flat background with $\omega=1$.
We introduce a fluctuation $h^{\mu\nu}$ around the flat metric $\eta^{\mu\nu}$ defined by
\begin{equation}
\tilde{g}^{\mu\nu}=\eta^{\mu\nu}+h^{\mu\nu},
\end{equation}
and then to the linear order we have
\begin{equation}
g_{\mu\nu}=\eta_{\mu\nu}-h_{\mu\nu}+\frac{1}{2}\eta_{\mu\nu}h+\cdots, \qquad \sqrt{-g}=1+\frac12 h+\cdots
\end{equation}
In what follows indices of the fields will be raised and lowered by using $\eta^{\mu\nu}$ and $\eta_{\mu\nu}$,
respectively. The quadratic terms in our total action are given by
\begin{eqnarray}
{\cal L}_{\rm UG}\Bigr|_{\rm quadr} &=&
{\cal L}_{N_{\FP}=0} + {\cal L}_{|N_{\FP}|=1} + {\cal L}_{|N_{\FP}|=2} +{\cal L}_{|N_{\FP}|=3},  \nn
{\cal L}_{N_{\FP}=0}
&=&   \frac14 h_{\mu\nu}\square h^{\mu\nu}+\frac12 (\pa_\nu h^{\mu\nu})^2
-\frac18 h\square h +\frac12 \lambda h
+ b_{\mu} (\pa_\la h^{\mu\la}-\pa^\mu S)
, \nn
{\cal L}_{|N_{\FP}|=1}&=&
\frac{i}{2} \varepsilon^{\mu\nu\rho\s} \bar c_\mu\square\pa_\nu C_{\rho\s}
-i \bar c_\mu\pa^\mu C_S + i C \pa_\mu\bar C^\mu+i \bar C^\mu\pa^\nu C_{\nu\mu}
, \nn
{\cal L}_{|N_{\FP}|=2}&=&
 -\bar D^\mu(\square D_\mu-\pa^\nu\pa_\mu D_\nu)
+ \bar D\, \pa^\mu D_\mu+\a \bar D D +\pa_\mu\bar D^\mu\cdot D , \nn
{\cal L}_{|N_{\FP}|=3}&=& i \bar T\, \square T.
\end{eqnarray}

We start with $N_{\FP}=0$ sector. The 2-point vertex $\Gamma^{(2)}_{N_{\FP}=0}$ in momentum space is
\begin{eqnarray}
\hspace{-1em}\Gamma^{(2)}_{N_{\FP}=0} =
&&\hs{-5} \bordermatrix{
           & h_{\rho\sigma}  & S & b_\rho& \la  \cr
h_{\mu\nu} &
\begin{matrix}
\hs{-10} -p^2\,\Bigl[\frac12 P^{(2)\mu\nu,\rho\sigma}-\frac1{12}d^{\mu\nu}d^{\rho\sigma}
\\
-\frac14\left(d^{\mu\nu}e^{\rho\sigma}+e^{\mu\nu}d^{\rho\s}\right)
-\frac34e^{\mu\nu}e^{\rho\sigma}\Bigr]
\end{matrix}
   & 0 & -i p^{(\mu} \eta^{\nu)\rho} & \frac12\eta^{\mu\nu}      \cr
S & 0 & 0 & ip^\rho& 0  \cr
b_\mu& i p^{(\rho} \eta^{\s)\mu}  & -ip^\mu & 0 & 0 \cr
\la & \frac12 \eta^{\rho\sigma}  & 0 & 0 &  0 \cr
},
\end{eqnarray}
by using the projection operators
\def\II{{I\kern-2.5ptI}}
\begin{eqnarray}
&&d_{\mu\nu} = \eta_{\mu\nu}- \frac{p_\mu p_\nu}{p^2}, \qquad
e_{\mu\nu}= \frac{p_\mu p_\nu}{p^2},  \\
&& P^{(2)}_{\mu\nu,\rho\sigma}= \frac12\left(
d_{\mu\rho}d_{\nu\sigma}+d_{\mu\sigma}d_{\nu\rho}-\frac23 d_{\mu\nu}d_{\rho\sigma}\right),
\end{eqnarray}
which satisfy
\begin{eqnarray}
&&p^\mu d_{\mu\nu}=0,\quad d_{\mu\nu}\eta^{\mu\nu}=3,\quad e_{\mu\nu}\eta^{\mu\nu}=1, \\
&&
d_{\mu\alpha}d^{\alpha\nu}=d_\mu{}^\nu,\quad e_{\mu\alpha}e^{\alpha\nu}=e_\mu{}^\nu,
\quad d_{\mu\alpha}e^{\alpha\nu}=0, \\
&& P^{(2)}_{\mu\nu,\alpha\beta}d^{\alpha\beta}= 0,\quad
P^{(2)}_{\mu\nu,\alpha\beta}e^{\alpha\beta}= 0,\quad
P^{(2)}_{\mu\nu,\alpha\beta}P^{(2) \alpha\beta,\rho\sigma}= P^{(2)}_{\mu\nu}{}^{\rho\sigma}.
\end{eqnarray}

We can straightforwardly compute the inverse of the matrix,
${\Gamma^{(2)}}^{-1}_{N_{\FP}=0}$:
{\small
\begin{eqnarray}
&&\hspace{-1em}{\Gamma^{(2)}}^{-1}_{N_{\FP}=0}=
\frac1{-p^2}\times
\nonumber \\
&& \hs{-8}\bordermatrix{
           & h_{\rho\sigma}  & S & b_\rho& \la  \cr
h_{\mu\nu} &
\begin{matrix}
\Bigl[ 2 P^{(2)}_{\mu\nu,\rho\sigma} - \frac13 d_{\mu\nu}d_{\rho\sigma}
\\
+\left(d_{\mu\nu}e_{\rho\sigma}+e_{\mu\nu}d_{\rho\sigma}\right)
-3 e_{\mu\nu}e_{\rho\sigma}\Bigr]
\end{matrix}
    & d_{\mu\nu}-3 e_{\mu\nu} & 2i p_{(\mu} d_{\nu)\rho} & -p^2 (d_{\mu\nu}-e_{\mu\nu})  \cr
S & d_{\rho\s}-3 e_{\rho\s} & -3 & -ip_\rho& p^2 \cr
b_\mu& -2i p_{(\rho} d_{\s)\mu} & ip_\mu& 0 & 0 \cr
\la & -p^2 ( d_{\rho\sigma}-e_{\rho\sigma})  & p^2 & 0 & 0 \cr
}.~~
\label{eq:GammaInNF0}
\end{eqnarray}
}
\indent
The 2-point vertex $\Gamma^{(2)}_{|N_{\FP}|=1}$ in momentum space is
\begin{eqnarray}
\Gamma^{(2)}_{|N_{\FP}|=1} =
\bordermatrix{
         &   C_{\rho\s} & C & C_S \cr
\bar c_\mu& \frac{1}{2} \varepsilon^{\mu\nu\rho\s} p^2 p_\nu& 0 & p^\mu\cr
\bar C_\mu&  -p^{[\rho}\eta^{\s]\mu} &  -p^\mu& 0 \cr
},
\end{eqnarray}
the inverse of which is given by
\begin{equation}
{\Gamma^{(2)}}^{-1}_{|N_{\FP}|=1}=
\frac1{-p^2}\times
\bordermatrix{
         &  \bar c_\rho& \bar C_\rho\cr
C_{\mu\nu} & \varepsilon_{\mu\nu\rho\la} p^\la/p^2 & 2p_{[\mu}\eta_{\nu]\rho}  \cr
C & 0 & p_\rho\cr
C_S & -p_\rho& 0 \cr
}.
\label{eq:GammaInNF1}
\end{equation}

The 2-point vertex $\Gamma^{(2)}_{|N_{\FP}|=2}$ in momentum space is
\begin{eqnarray}
\Gamma^{(2)}_{|N_{\FP}|=2} =
\bordermatrix{
         &   D_\rho& D \cr
\bar D^\mu& p^2 d_\mu{}^\rho& -ip_\mu\cr
\bar D  &  ip^\rho&  \a  \cr
},
\end{eqnarray}
the inverse of which is given by
\begin{equation}
{\Gamma^{(2)}}^{-1}_{|N_{\FP}|=2}=
\frac1{-p^2}\times
\bordermatrix{
         & \bar D^\rho& \bar D \cr
D_\mu& \left[-\d_\mu{}^\rho+(\a+1) p_\mu p^\rho/p^2
\right] & ip_\mu\cr
D &  -ip^\rho& 0  \cr
}.
\label{eq:GammaInNF2}
\end{equation}

Finally the 2-point vertex $\Gamma^{(2)}_{|N_{\FP}|=3}$ for $T$ and $\bar{T}$ in momentum space is given by
\begin{eqnarray}
\Gamma^{(2)}_{|N_{\FP}|=3} = -ip^2,
\end{eqnarray}
and the inverse of which is given by
\begin{equation}
{\Gamma^{(2)}}^{-1}_{|N_{\FP}|=3}=\frac{i}{p^2}.
\label{eq:GammaInNF3}
\end{equation}

Thus we have confirmed that the propagator may be obtained and the system is fully gauge fixed.

\subsection{Equations of motion at linear order}
\label{eom}

Let us denote the total action as $S$. The EOMs to linear order are given as follows:
for   $\alpha=-1$,
\begin{align}
\hs{-20}
\hbox{$N_\FP=0$ sector}\ & \nn
\frac{\delta S_{\rm UG}}{\delta\lambda} :\quad & h(\equiv\eta_{\mu\nu}h^{\mu\nu})=0,
\label{eq:N01} \\
\frac{\delta S_{\rm UG}}{\delta h^{\mu\nu}} :\quad & \frac{1}{2}\square h_{\mu\nu}-\partial_{(\mu}(h_{\nu)}
+b_{\nu)})+\frac{1}{2}\lambda\eta_{\mu\nu}=0,
\label{eq:N02}\\
\frac{\delta S_{\rm UG}}{\delta b^\mu} :\quad & \pa^\nu h_{\mu\nu}-\pa_\mu S=0,
\label{eq:N03}\\
\frac{\delta S_{\rm UG}}{\delta S}:\quad & \pa^\mu b_\mu=0,
\label{eq:N04}
\end{align}
\begin{align}
\hs{-25}
\hbox{$N_\FP=\pm1$ sector}\ &  \nn
\frac{\delta S_{\rm UG}}{\delta\bar{c}^{\mu}} :\quad
& \frac{1}{2} \varepsilon^{\mu\nu\rho\s} \square\pa_{\nu} C_{\rho\s}
 - \pa^{\mu}C_S=0,
\label{eq:N11} \\
\frac{\delta S_{\rm UG}}{\delta C_{\mu\nu}} :\quad
& \frac12 \varepsilon^{\mu\nu\rho\s} \square\pa_\rho\bar c_\s
 - \pa^{[\mu}\bar C^{\nu]}=0,
\label{eq:N12} \\[.5ex]
\frac{\delta S_{\rm UG}}{\delta\bar{C}^{\mu}} :\quad
& \pa^\nu C_{\nu\mu} + \pa_\mu C=0,
\label{eq:N13} \\
\frac{\delta S_{\rm UG}}{\delta C} :\quad
& \pa^\mu\bar C_\mu=0,
\label{eq:N14} \\
\frac{\delta S_{\rm UG}}{\delta C_S} :\quad
& \pa^\mu\bar c_\mu=0,
\label{eq:N15}
\end{align}
\vs{-10}
\begin{align}
\hbox{$N_\FP=\pm2$ sector}\  & \qquad \nn
\frac{\delta S_{\rm UG}}{\delta\bar{D}}, \ \frac{\delta S_{\rm UG}}{\delta D} :\quad
& \partial^{\mu}D_{\mu}=D,\quad \partial_{\mu}\bar{D}^{\mu}=\bar{D},\qquad\qquad
\label{eq:N21} \\
\frac{\delta S_{\rm UG}}{\delta\bar{D}^{\mu}},\ \frac{\delta S_{\rm UG}}{\delta D_{\mu}} :\quad
& \square D_{\mu}=0,\quad \square\bar{D}^{\mu}=0,
\label{eq:N22} \\[.5ex]
\hbox{$N_\FP=\pm3$ sector}\ &  \nn
\frac{\delta S_{\rm UG}}{\delta\bar{T}},\ \frac{\delta S_{\rm UG}}{\delta T} :\quad
 & \square T=0,\quad  \square\bar{T}=0 .
\label{eq:N3}
\end{align}
where $h_\mu\equiv\partial^\nu h_{\mu\nu}$.
Note also that Eqs.~(\ref{eq:N02}), (\ref{eq:N12}) and \eqref{eq:N22} are already
simplified by their preceding equations.

Taking the trace of \p{eq:N02} and using \p{eq:N01} and \p{eq:N04}, we find
\bea
2\la=\pa^\mu\pa^\nu h_{\mu\nu}.
\label{eq:la}
\eea
The divergence of the gravity equation (\ref{eq:N02}), combined with \p{eq:la} and \p{eq:N04}, yields
\bea
\square b_\mu=-\pa_\mu\la,
\label{eq:bmu}
\eea
which, together with \eqref{eq:N04}, gives
\bea
\square\la=0.
\eea
Equation~\eqref{eq:N03} gives
\bea
-\square S + 2\la =0\,.
\label{eq:BoxS}
\eea
Equations \eqref{eq:bmu} and \eqref{eq:BoxS}) imply that $b_\mu$ and $S$ fields satisfy dipole equations
$\square^2 b_\mu=0$ and $\square^2 S=0$, and their dipole parts are supplied by the simple pole $\la$ field.
Applying $\square$ to the gravity field equation (\ref{eq:N02}) and using
Eqs.~(\ref{eq:bmu}) and (\ref{eq:BoxS}), we get
\bea
\frac12\square^2 h_{\mu\nu} - \partial_\mu\partial_\nu\la =0.
\label{eq:Box2hmunu}
\eea
So, we see that $\tlh_{\mu\nu}$ field is now a tripole field, and the tripole part
is supplied by the simple pole $\la$ field and the dipole parts are
supplied by simple pole parts of $S$ and $b_\mu$.

For the $N_{\rm FP}\ne 0$ ghost sector, we find from \eqref{eq:N11}
\bea
\square C_S=0.
\eea
The dual of \eqref{eq:N11} gives
\bea
\square^2 C_{\mu\nu}=0.
\eea
Taking the $\pa_\nu$ divergence of \eqref{eq:N12} and using \eqref{eq:N14} gives
\bea
\square\bar C^\mu=0.
\eea
The dual of \eqref{eq:N12} yields
\bea
\square^2 \bar c_\mu=0.
\eea
Equations \eqref{eq:N13}, (\ref{eq:N22}) and (\ref{eq:N3}) show that
all the other ghost fields are of simple pole:
\bea
\square C=
\square D= \square \bar{D}=
\square D_\mu= \square \bar{D}^\mu=
\square T= \square \bar{T} =
0.
\eea
There are no tripole ghost fields, in contrast to our paper I, though there are in the graviton
fluctuation. This is one of the simplifications that our new formulation brings in.

\section{Identifying the independent fields}
\label{counting}

We have not only the usual simple pole fields but also dipole and tripole
fields in this system. So the number of independent particle modes are
different from that of independent fields. To avoid the complication of
decomposing the multipole fields into simple pole modes, we also adopt
here, as in the previous paper I, the 4-dimensional Fourier
expansion\cite{Nakanishi:1966zz} of the fields
\begin{equation}
\phi(x) = \frac1{\sqrt{(2\pi)^3}}\int d^4p\, \theta(p^0)\left[
\phi(p)e^{ipx} + \phi^\dagger(p)e^{-ipx} \right]
\end{equation}
to define annihilation and creation operators $\phi(p)$ and
$\phi^\dagger(p)$ for such general multipole fields.
We analyze independent 4-dimensional Fourier modes for each ghost
number $N_\FP$ sector successively,
in the Lorentz frame in which the 3-momentum $\mbf{p}$ is along $x^3$
axis:
\begin{equation}
p^\mu= \bigl(p^0, 0, 0, p^3\bigr), \quad \text{i.e.,}\quad
p^i=0\ (i=1,2), \
p^3=:|p|>0.
\label{eq:frame}
\end{equation}
Note that if the field $\phi$ is a massless simple pole field $\phi(p)\propto\delta(p^2)$,
this reads
\begin{equation}
p^\mu\,\phi(p^2)=  \bigl( |p|, 0, 0, |p|  \bigr) \,\phi(p^2),
 \quad \text{i.e.,} \quad
p^0=p^3=|p|\ .
\label{eq:onshellp}
\end{equation}

We will show in the following analysis that the independent fields in
each ghost number sector are given by the list in Table~\ref{table:1}.
\begin{table}[htb]
\caption{List of independent fields. $i$ denotes transverse directions 1 and 2.}
\label{table:1}
\begin{center}
\begin{tabular}{|r|ccc|} \hline\hline
$N_\FP=0$
 & \Tspan{$ 
   h_{\T1},\ h_{\T2};\quad \chi^0,\ \chi^i $}
 &;& \Tspan{$ 
   b_0,\ b_i $} \\ \hline
$|N_\FP|=1$    & \Tspan{$ 
   {C}_{0i},\ {C}_{12},\ C $}
  &;& \Tspan{$ \bar{c}_0,\ \bar{c}_i,\ \bar{C}^0 $} \\ \hline
$|N_\FP|=2$    & \Tspan{$ D_0,\ D_i,\ D $}
  &;&  \Tspan{$ \bar{D}^0,\ \bar{D}^i,\ \bar{D} $} \\ \hline
$|N_\FP|=3$      & $T$ &;& \Tspan{$\bar{T}$}              \\ \hline
\end{tabular}
\end{center}
\end{table}

\subsection{$N_\FP=0$ sector}
\label{counting.1}

We begin with the fields with ghost number $N_\FP=0$.
We have 10 component gravity $h_{\mu\nu}$ field, 1 scalar field $S$,
1 unimodular multiplier field $\lambda$, plus a 4 component vector NL
field $b_\mu$; thus, $10+1+1+4=16$ component fields in all.
Let us count/identify the independent fields among them,
by using the EOMs~(\ref{eq:N01}) -- (\ref{eq:N04}).

These EOMs (\ref{eq:N01}) -- (\ref{eq:N04}) in this
$N_\FP=0$ sector take exactly the same forms as
those in GR theory in unimodular gauge, if we identify the unimodular
NL field $b$ there with the present unimodular multiplier $\lambda$. The same
counting there also applies here.

The 10 component $h_{\mu\nu}$ is subject
to the 1 traceless condition (\ref{eq:N01}) and the 4-vector
de Donder gauge condition (\ref{eq:N03}), so has $10-1-4=5$ independent
fields, as which we can take the same fields as in GR case.
First, we have two BRST invariant simple-pole (hence physical) fields
\begin{align}
&h_{\T1}(p):= \half \bigl(h^{11} - h^{22}\bigr)(p),  \qquad
h_{\T2}(p):= h^{12}(p).
\end{align}
These transverse modes are BRST invariant since the BRST transformation
of $h^{\mu\nu}(p)$ at linearized level is given by
\begin{equation}
\delta_\B h^{\mu\nu}(p) = -i p^\mu c^\nu_\T(p)-ip^\nu c^\mu_\T(p),
\end{equation}
while the transverse
momentum components $p^i\ (i=1,2)$ vanish by definition. Simple-pole
property $\square h_{\T j}(p)=0$ also follows from the EOM (\ref{eq:N02}) and $p^i=0$.
In addition to these two, we have a transverse vector field (hence possessing
3 independent components):
\begin{align}
&\chi^0(p) := \frac1{2p^0}\Bigl(h^{00}-\frac12\bigl(h^{11} +h^{22}\bigr)\Bigr)(p)
= \frac1{2p^0}\frac12\bigl(h^{00} + h^{33}\bigr)(p), \nn
&\chi^i(p) := \frac1{p^0}h^{0i}(p), \ \ (i=1,2), \nn
&\chi^3(p) := \frac1{2p^3}\Bigl(h^{33}-\frac12\bigl(h^{11}
+h^{22}\bigr)\Bigr)(p)
= \frac1{2p^3}\frac12\bigl(h^{00} + h^{33}\bigr)(p),
\label{eq:chimu}
\end{align}
satisfying transversality $p_\mu\chi^\mu(p)=p_0\chi^0(p)+p_3\chi^3(p)=0$.
So we can forget the redundant component $\chi^3(p)$ henceforth.
The second equality for the expression $\chi^0(p)$ (or $\chi^3(p)$) follows
from the tracelessness Eq.~(\ref{eq:N01}), $h:=\eta_{\mu\nu}h^{\mu\nu}=0$,
\begin{equation}
\left(h^{11}+h^{22}\right)(p) =
\left(h^{00}-h^{33}\right)(p).
\label{eq:h11+h22}
\end{equation}
This $\chi^\mu(p)$ field has a very simple BRST transformation property
\begin{equation}
\delta_\B \chi^\mu(p) = -i c^\mu_\T(p).
\label{eq:BRSchimu}
\end{equation}

Next, the gauge fixing NL field $b_\mu$ is subject to the transversality
(\ref{eq:N04}), $p^0b_0(p)+p^3b_3(p)=0$, so we can take 3 fields
$b_0$ and $b_i(p)$
$(i=1,2)$ as its independent components.
These 5 components $h_{\T i}$, $\chi^0$ and $\chi^i$ from $h_{\mu\nu}$ and
3 components $b_0$ and $b_i$ give the all of the independent fields
listed in the $N_\FP=0$ sector in Table~\ref{table:1}.

The other fields $\lambda$ and $S$ as well as the other 5 dependent components
in $h_{\mu\nu}$
\begin{equation}
\bigl(h^{11}+h^{22}\bigr)(p),\quad
\bigl(h^{00}-h^{33}\bigr)(p),\quad
h^{3i}(p),\quad h^{03}(p).
\label{eq:dependentHmunu}
\end{equation}
can be shown to be explicitly expressed by the above $5+3$ independent fields
by using EOMs (\ref{eq:N01}) -- (\ref{eq:N04}) as follows.

In Eq.~(\ref{eq:bmu}) which we already derived from those EOMs,
the unimodular multiplier $\lambda$ was identified with the dipole part
of the gauge-fixing multiplier (NL) field $b_0$:
\begin{equation}
\lambda(p)
=-i\frac1{p^0}\square b_0(p)\,. 
\label{eq:lambda=Boxb0}
\end{equation}
This is an important relation showing that the unimodular multiplier field
$\lambda$ becomes the BRST daughter field, hence a member of a BRST quartet.
As for the $S$ field, we can rewrite Eq.~(\ref{eq:N02}) into the following form
by substituting $h_\mu=\partial_\mu S$ of Eq.~(\ref{eq:N03}):
\begin{equation}
\partial^\mu\partial^\nu S = \frac12\square h^{\mu\nu}-\partial^{(\mu}b^{\nu)}+\frac12\lambda\eta^{\mu\nu}.
\end{equation}
Adding two equations with indices $(\mu,\nu)=(0,0)$ and $(3,3)$ and
dividing by $-(p_0^2+p_3^2)$, we obtain
\begin{equation}
S(p) = -\frac{2p^0}{p_0^2+p_3^2}\bigl(\square\chi^0-ib^0\bigr)(p).
\label{eq:S}
\end{equation}

Now the 5 dependent components in $h_{\mu\nu}$ in Eq.~(\ref{eq:dependentHmunu}).
The first dependent field $(h^{11}+h^{22})(p)$
is already given by the second field $(h^{00}-h^{33})(p)$ in
Eq.~(\ref{eq:h11+h22}). The latter one is shown in Eq.~(4.43) in
the previous paper I to be
\begin{align}
&\left(h^{00}-h^{33}\right)(p) =  \frac{4ip^0}{p_0^2+p_3^2}\,b_0(p)\,,
\end{align}
which is actually the same equation as Eq.~(\ref{eq:S}) if we substitute
$-(p_0^2+p_3^2)S(p)= i(p^0h^0+p^3h^3)(p) = p_0^2h^{00}(p)-p_3^2h^{33}(p)$
following from $\partial^\nu\partial^\mu S=\partial^\nu h^\mu=\partial^\nu\partial_\lambda h^{\mu\lambda}$.
The rest two components $h^{3i}$ and $h^{03}$ follow from the de Donder gauge
condition (\ref{eq:N03}), $\partial_\nu h^{\nu\mu}=\partial^\mu S$; the $\mu=i$ ($i=1,2$)
component gives
\begin{equation}
h^{3i}(p) 
= \frac{p_0^{\,2}}{p^3}\,\chi^i(p)\,,
\end{equation}
and the vanishing difference $0=\partial^{[\nu}\partial^{\mu]}S = \partial^{[\nu}\partial_\lambda h^{\mu]\lambda}$
with $(\mu,\nu)=(0,3)$ gives Eq.~(4.41) in I:
\begin{equation}
h^{03}(p) = \frac{4p_0^2p^3}{p_0^2+p_3^2}\chi^0(p).
\label{eq:H03}
\end{equation}

\subsection{$N_\FP\not=0$ sectors}

In $N_\FP=+1$ ghost sector we have 6 component $C_{\mu\nu}$ and 2 scalars,
$C_S$ and $C$,  so 8 components in all.
In $N_\FP=-1$ antighost sector we have two 4-component vectors
$\bar c_\mu$ and $\bar{C}^\mu$, so also 8 components in all.

Begin with the $N_\FP=+1$ ghost sector. The constraint equation (\ref{eq:N13}),
$\partial^\nu C_{\nu\mu}+\partial_\mu C=0$, takes the same form as that in the previous paper I,
which implied that only 3 components $C_{0i}$ and $C_{12}$ are independent
among 6 $C_{\mu\nu}$ if the scalar $C$ is chosen as another independent field;
the other 3 components are expressed by these and $C$
as the Eq.~(4.52) in I with $\bar{C}^{\mu\nu}$ and $\bar{C}$ there replaced by
$C_{\mu\nu}$ and $C$:
\begin{equation}
C_{03}= -C(p), \qquad C_{3i}=-\frac{p^0}{p^3}C_{0i}(p),
\quad (i=1,2).
\label{eq:C03-3i}
\end{equation}
These $3+1=4$ fields $C_{0i}$, $C_{12}$ and $C$ are
the all of the independent fields listed in the $N_\FP=1$ sector in Table~\ref{table:1}.
The other remaining field $C_S$ is indeed expressed by the dipole part
of the independent field $C_{12}$; the $\mu=0$ component of EOM (\ref{eq:N11})
gives
\begin{equation}
C_S(p) = \frac1{2p^0} \varepsilon^{03ij}\square p_3 C_{ij}(p)
= \square C_{12}(p),
\end{equation}
where in the second equality, $p_3=p^0$ has been used since $\square C_{12}(p)$ is a massless
simple pole field.

Next consider the $N_\FP=-1$ antighost sector, consisting of two vectors
$\bar c_\mu$ and $\bar{C}^\mu$.
Equations~(\ref{eq:N15}) and (\ref{eq:N14}) show that these vectors are both transversal:
\begin{align}
p^0\bar{c}_0 + p^3\bar{c}_3 =0 \ &\rightarrow\ \bar{c}_3 = -\frac{p^0}{p^3} \bar{c}_0, \\
p_0\bar{C}^0 + p_3\bar{C}^3 =0 \ &\rightarrow\ \bar{C}^3 = \frac{p^0}{p^3} \bar{C}^0 =
\bar{C}^0\,,
\end{align}
where $p^0=p^3$ has been used in front of the simple pole field $\bar{C}^\mu$.
Moreover, Eq.~(\ref{eq:N12}) gives constraint relations between
$\bar{c}_\mu$ and $\bar{C}^\nu$: noting that $\square\bar{c}_\mu$ is also a simple
pole field,
\begin{align}
\text{($\mu=0$, $\nu=3$)} &:~  p^0 \bar{C}^3-p^3\bar{C}^0 = 0 \ \rightarrow\ \bar{C}^0=\bar{C}^3, \label{eq:barC03}\\
\text{($\mu=0$ or 3, $\nu=i$)} &:~  \varepsilon^{ij}\square p_3 \bar{c}_j(p) + p^0\bar{C}^i=0
\ \rightarrow\
\bar{C}^i(p) = -\varepsilon^{ij}\square\bar{c}_j(p) \nn
&\hspace{13em} (i,j = 1,2,\ \varepsilon^{ij}=-\varepsilon^{ji}),
\label{eq:barci}\\
\text{($\mu=1$, $\nu=2$)} &:~
\square p_0\bar{c}_3 - \square p_3\bar{c}_0 = 0 \ \rightarrow\
\square\bar{c}_3 = - \square\bar{c}_0 \,.
\label{eq:barc03}
\end{align}
Equations~(\ref{eq:barC03}) and (\ref{eq:barc03}) merely give identical relations
with the transversality of vectors $\bar{c}_\mu$ and $\bar{C}^\nu$.
Only the Eq.~(\ref{eq:barci}) implies new relations that
$\bar{C}^i(p)$ are given by the dipole part of $\varepsilon^{ij}\bar{c}_j(p)$.
We thus have shown that $3+1=4$ fields $\bar{c}_0$, $\bar{c}_i$ and
$\bar{C}^0$ are all of the independent fields in $N_\FP=-1$
antighost sector as listed in Table~\ref{table:1}.

Finally, the EOMs (\ref{eq:N21}) -- (\ref{eq:N3}) in the $|N_\FP|=2,3$
sectors are trivial. So we can immediately see that the independent
fields in these sectors can be chosen as written in the list of Table~\ref{table:1}.

\section{Identifying BRST quartets}
\label{ident}

Now that all the independent fields are listed up in Table~\ref{table:1},
we can show that all of them
other than the two physical transverse modes $h_{\T1}$ and
$h_{\T2}$, fall into BRST quartets, decoupling properly from the physical sector.
In order to do so, we recall the BRST transformation at linearized level
(which is also the BRST transformation of the asymptotic fields under the
perturbative assumption):
\begin{align}
\delta_\B \lambda&= 0,
\label{eq:BRSlambda}\\
\delta_\B h^{\mu\nu} &= -2\partial_{\phantom{T}}^{(\mu}c_{\T}^{\nu)}
= \partial^{(\mu}\varepsilon^{\nu)\tau\rho\sigma}\partial_\tau C_{\rho\sigma},
\label{eq:BRShmunu}\\
\delta_\B C_{\mu\nu} &= i (\partial_\mu D_\nu-\partial_\nu D_\mu),
\label{eq:BRSCmunu}\\
\delta_\B S&= C_S, \hspace{1em}\qquad  \delta_\B C_S=0,
\label{eq:BRSS} \\
\delta_\B \bar c_\mu &= i b_\mu, \hspace{1em}\qquad  \delta_\B b_\mu=0,
\label{eq:BRSbarcmu}\\
\delta_\B \bar D^\mu&= \bar C^\mu, \hspace{1em}\qquad  \delta_\B \bar C^\mu=0,
\label{eq:BRSbardmu}\\
\delta_\B D_\mu&=  \partial_\mu T,\hspace{.7em}\qquad \d_\B T=0,
\label{eq:BRSdmu}\\
\delta_\B \bar T &=  i \bar D, \hspace{1.1em}\qquad  \delta_\B \bar D=0,
\label{eq:BRSbart}\\
\delta_\B C &=  iD, \hspace{1.1em}\qquad \delta_\B D=0.
\label{eq:BRSC}
\end{align}

The BRST quartet is generally a pair of the BRST doublets which satisfy the
properties schematically drawn as~\cite{KO1977}
\begin{equation}
\xymatrix{
 A(p)\ \ar@{->}[r]^{\delta_{\text{B}}}\ar@{<.>}[rrd] 
                   & \ C(p) \ar@{<..>}[d]_{\text{inner-product}} &  \\
   &\ \bar C(p)\  \ar[r]_{\delta_{\text{B}}} & \ iB(p) .
 }
\label{eq:BRSquartet}
\end{equation}
We denote this BRST quartet described by this scheme by using
the same notation as used in I simply as
\begin{equation}
\bigl( A(p) \rightarrow C(p);\ \bar C(p) \rightarrow iB(p) \bigr),
\label{eq:QuartetNotation}
\end{equation}
This scheme means that
a pair of BRST doublets satisfying (assuming $A(p)$ a boson),
\begin{align}
\delta_{\B}  A(p) &= [iQ_\B,\,A(p)] = C(p), \nn
\delta_{\B}  \bar C(p) &= \{iQ_\B,\,\bar{C}(p) \} = iB(p),
\end{align}
have nonvanishing inner-product with each other:
\begin{equation}
\langle0| \bar{C}(p) C^\dagger(q) |0\rangle=
\langle0| \bar{C}(p)iQ_\B A^\dagger(q) |0\rangle= i\langle0| B(p) A^\dagger(q) |0\rangle
\propto\delta^4(p-q) \not=0,
\label{eq:Innerproduct1}
\end{equation}
or, equivalently, in terms of commutation relation,
\begin{align}
0&=\bigl\{ iQ_\B,\, [\bar{C}(p),\, A^\dagger(q)] \bigr\} =
[i B(p),\, A^\dagger(q)] - \{ \bar{C}(p),\, C^\dagger(q) \}  \nn
&\rightarrow\
[i B(p),\, A^\dagger(q)] = \{ \bar{C}(p),\, C^\dagger(q) \}
\propto\delta^4(p-q) \not=0.
\label{eq:Innerproduct2}
\end{align}

We should note that the existence of nonvanishing inner-products/commutators
can also be
judged from the nonvanishing propagators. As explained in I, generally,
the expressions of commutation relations (CR) and the propagators for the
free fields $\phi_i$ have the following exact correspondence:
\begin{equation}
\begin{array}{ccc}
\text{propagator}\ \langle\phi_i\, \phi_j \rangle&&  \text{CR}\  [\phi_i(p),\, \phi^\dagger_j(q)]
\\[.1ex]
\hline
\phantom{\bigg|}
\displaystyle \frac1{i} \Bigl[ \frac1{p^2},\  \frac1{p^4},\
\frac1{p^6} \Bigr] 
&\leftrightarrow&
\Bigl[ \delta(p^2),\  -\delta'(p^2),\ \frac12\delta''(p^2) \Bigr] \theta(p^0)\delta^4(p-q),
\end{array},
\label{eq:translation}
\end{equation}
where three terms on both sides correspond to the terms of simple-pole,
dipole and tripole parts, respectively. This correspondence also holds
for the fermion fields if the commutators are understood to be anticommutators.
We use an abbreviated notation for the propagator in momentum space:
\begin{equation}
\langle\,\phi_i\ \phi_j\,\rangle_p = \int d^4x\, e^{-ipx} \VEV{\T \phi_i(x)\,\phi_j(0)},
\end{equation}
which is identical with $i$ times the inverse of the 2-point vertex,
$i{\Gamma^{(2)}}^{-1}_{ij}$, so can be read directly from the results given in sect.~3.

Now let us start the task to identify the BRST quartets.
Among the fields in Table~\ref{table:1}, we can list all the boson fields
which are not BRST invariant. Then they must be BRST parents
of certain BRST doublets, which can be read
from the BRST transformation Eqs.~(\ref{eq:BRSlambda}) -- (\ref{eq:BRSC}).
Thus we find all the BRST doublets possessing
boson parent as
\begin{align}
\delta_\B \chi^0(p) &= -ic^0_\T(p) = -p_3C_{12}(p),
\label{eq:Db-1}\\
\delta_\B \chi^i(p) &= -ic^i_\T(p) = \varepsilon^{ij}\bigl(p_3C_{0j}-p_0C_{3j}\bigr)(p)
= - \varepsilon^{ij}\frac1{p^3}\square C_{0j}(p),
\label{eq:Db-2}\\
\delta_\B D^0(p) &= ip^0 T(p),
\label{eq:Db-3}\\
\delta_\B \bar{D}^0(p) &= \bar{C}^0(p),
\label{eq:Db-4}\\
\delta_\B \bar{D}^i(p) &= \bar{C}^i(p)= -\varepsilon^{ij}\square\bar{c}_j(p),
\label{eq:Db-5}
\end{align}
where in Eq.~(\ref{eq:Db-2}), use has been made of Eq.~(\ref{eq:C03-3i})
to rewrite $C_{3j}(p)$ as $-(p^0/p^3)C_{0j}(p)$ in the last equality.
We have used Eq.~(\ref{eq:barci}) in the last line to rewrite $\bar{C}^i$
which does not appear in Table~\ref{table:1} as independent fields.
In the same way, all the BRST doublets possessing fermion parent can be
found as
\begin{align}
\delta_\B C_{0i}(p) &= -p_0 D_i(p),
\label{eq:Df-1}\\
\delta_\B C(p) &= iD(p),
\label{eq:Df-2}\\
\delta_\B \bar{c}_0(p) &= ib_0(p),
\label{eq:Df-3}\\
\delta_\B \bar{c}_i(p) &= ib_i(p),
\label{eq:Df-4}\\
\delta_\B \bar{T}^i(p) &= i\bar{D}(p).
\label{eq:Df-5}
\end{align}

At this stage we notice that all the independent fields in Table~\ref{table:1}
other than the physical transverse graviton modes already appear in these
BRST doublets, as the parent or daughter fields.
Since the BRST doublets are known to necessarily form
BRST quartets which essentially decouple from physical subspace, one is
tempted to conclude that the unitarity proof is completed.
We must, however, be very careful about the dipole and tripole fields
appearing in the BRST parent position.
For instance, if $A$ is a dipole and $B$ is a simple pole for the BRST doublet
$\delta_\B A = iB$, then this implies
$\delta_\B \square A = 0$. That is, the simple pole field $\square A$, representing the dipole
part mode in the $A$ field is BRST invariant. So $\square A$ must appear somewhere
at BRST daughter position in the full list of the BRST doublets.
Otherwise, the unphysical $\square A$ mode becomes a BRST singlet appearing in the
physical subspace and violates unitarity.

All the multipole fields which appear in the parent position in these BRST
doublets (\ref{eq:Db-1}) -- (\ref{eq:Df-5})
are: $\chi^0,\ \chi^i,\ C_{0i},\ \bar{c}_0,\ \bar{c}_i$.
Let us examine the BRST doublets possessing these parent fields in turn.

First is the doublet (\ref{eq:Db-1}), $\delta_\B\chi^0(p)=-p_3C_{12}(p)$.
The parent $\chi^0(p)$ is
tripole while the daughter $C_{12}(p)$ is dipole, meaning that the tripole
mode part $\square^2\chi^0(p)$ could be a dangerous BRST singlet. But
Eq.~(\ref{eq:Box2hmunu}), $(1/2)\square^2h_{\mu\nu}=\partial_\mu\partial_\nu\lambda$, tells us
$\square^2\chi^0(p)\propto\lambda(p)$ and the unimodular multiplier field $\lambda(p)$ was
already noted in Eq.~(\ref{eq:lambda=Boxb0}) to be a safe BRST
daughter field $\propto\square b_0(p)$ and it indeed appears in the daughter
position of the BRST doublet (\ref{eq:Df-3}).
Note that this BRST doublet is a safe doublet, in which both
the parent $\bar{c}_0(p)$ and the daughter $b_0(p)$
are dipole fields, implying that it actually represents two simple-pole
BRST doublets:
\begin{equation}
\delta_\B \hat{\bar{c}}_0(p) = i\hat{b}_0(p) \quad \text{and}\quad
\delta_\B \square\bar{c}_0(p) = i\square b_0(p),
\label{eq:Df-3_2}
\end{equation}
where the hat symbol $\hat\phi$ means the simple-pole part in the multipole field.
Here we emphasize that the second equation of (\ref{eq:Df-3_2}) means that
the multiplier field $\la$ imposing the unimodular constraint becomes
the BRST daughter $\square b_0=ip^0\lambda$ and has associated
ghost and antighost. This is the key how we get the correct number of
remaining physical dofs.

Second is the doublet (\ref{eq:Db-2}),
$\delta_\B\chi^i(p)=- \varepsilon^{ij}\square C_{0j}(p)/p^3$. Since
the parent $\chi^i(p)$ is a dipole field while $\square C_{0j}(p)$
is a simple-pole field, the dipole part $\square\chi^i(p)$ is a potentially
dangerous BRST singlet. But
EOMs (\ref{eq:N02}) and (\ref{eq:N03}) lead to
$\square\chi^i(p)=\square h^{0i}(p)/p^0= i b^i(p)$, which safely
appears in the daughter position of the doublet (\ref{eq:Df-4}).

Third is the doublet (\ref{eq:Df-1}), in which the parent $C_{0i}(p)$
is dipole while the daughter $D_i(p)$ is simple-pole.
So $\square C_{0i}(p)$ is a potentially dangerous BRST singlet but it
actually appears in the daughter position of the doublet (\ref{eq:Db-2}).

Lastly are the doublets (\ref{eq:Df-3}) and (\ref{eq:Df-4});
the former doublet (\ref{eq:Df-3}) was already noted in advance
to represent a safe and double BRST doublets in Eq.~(\ref{eq:Df-3_2}).
In the latter doublet (\ref{eq:Df-4}), the parent $\bar{c}_i(p)$ is
dipole while $b_i(p)$ is simple-pole. But the potentially dangerous
BRST singlet $\square\bar{c}_i(p)$ just
appears in the daughter position of the doublet (\ref{eq:Db-5}).

We thus have finished to show that the list of the BRST doublets in
Eqs.~(\ref{eq:Db-1}) -- (\ref{eq:Df-5}) is complete; that is, it
contains all the independent modes in our UG theory other than the
physical transverse modes and they all fall into BRST doublets
decoupling from the physical sector.

For the unitarity proof, this is enough. For completeness, however, we
identify how those BRST doublets form BRST quartets. The task for doing so
is only to find for each BRST doublet with
boson parent in (\ref{eq:Db-1}) -- (\ref{eq:Db-5})
a partner BRST doublet with fermion parent from (\ref{eq:Df-1}) --
(\ref{eq:Df-5}) which has nonvanishing inner-product with each other.
As explained before, the existence of nonvanishing inner-products can be
easily found by examining the nonvanishing propagators given in sect.~3.
In this way, we can identify the following BRST quartets:
\begin{align}
&\bigl(\ \chi^0(p) \ \rightarrow\ -p^3C_{12}(p);\
\ \bar{c}_0(p) \ \rightarrow\ i b_0(p) \ \bigr),
\label{eq:FirstQuartet}\\
&\bigl(\ \chi^i(p) \ \rightarrow\  -(1/p^3)\varepsilon^{ij}\square C_{0j}(p) ;\
\ \bar{c}_i(p) \ \rightarrow\ i b_i(p)\ \bigr),  \quad (i=1,2),
\label{eq:SecondQuartet}\\
&\bigl(\ D^0(p) \ \rightarrow\ ip^0 T(p) ;\
\bar{T}(p) \ \rightarrow\ i\bar{D}(p) \bigr),
\label{eq:3rdQuartet}\\
&\bigl(\ \bar{D}^0(p) \ \rightarrow\ \bar{C}^0(p) ;\
C(p) \ \rightarrow\ iD(p) \bigr),
\label{eq:4thQuartet}\\
&\bigl(\ \bar{D}^i(p) \ \rightarrow\ -\varepsilon^{ij}\square\bar{c}_j(p) ;\
C_{0i}(p) \ \rightarrow\ -p_0 D_i(p) \bigr).
\label{eq:5thQuartet}
\end{align}

The relevant propagators showing the existence of nonvanishing inner-products
between the pair doublets in these BRST quartets are the following:
\begin{align}
&\langle\,ib_0\ \chi^0\,\rangle_p
= p^3\langle\,\bar{c}_0\ C_{12}\,\rangle_p
= i \frac{p_3^2}{(p^2)^2},
\label{eq:Prop-1}\\
&\langle\,ib_i\ \chi^i\,\rangle_p
= \frac1{p^3}\langle\,\bar{c}_i\ \varepsilon^{ij}\square C_{0j}\,\rangle_p
= i\frac1{p^2},
\label{eq:Prop-2}\\
&\langle\,i\bar{D}\ D^0\,\rangle_p = -p^0\langle\,\bar{T}\ iT\,\rangle_p
= -i\frac{p^0}{p^2},
\label{eq:Prop-3}\\
&\langle\,iD\ \bar{D}^0\,\rangle_p=\langle\,C\ \bar{C}^0\,\rangle_p
= -i\frac{p^0}{p^2},
\label{eq:Prop-4}\\
& p^0\langle\,D_i\ \bar{D}^i\,\rangle_p
= - \langle\,C_{0i}\ \varepsilon^{ij}\square\bar{c}_j\,\rangle_p
= i\frac{p^0}{p^2}.
\label{eq:Prop-5}
\end{align}
Here we note that the propagators in the first line are of dipole.
It implies that the first BRST quartet (\ref{eq:FirstQuartet}) in fact
represents two BRST quartets each consisting of simple-pole fields:
\begin{align}
&\bigl(\ \square\chi^0(p) \ \rightarrow\ -p^3\square C_{12}(p);\
\ \hat{\bar{c}}_0(p) \ \rightarrow\ i \hat{b}_0(p) \ \bigr),
\label{FirstQuartet-1}\\
&\bigl(\ \hat{\chi}^0(p) \ \rightarrow\ -p^3\hat{C}_{12}(p);\
\ \square\bar{c}_0(p) \ \rightarrow\ i \square b_0(p) \ \bigr),
\label{FirstQuartet-2}
\end{align}
where hat means the simple-pole part. This structure can be understood
if we note the following relevant propagators of simple-pole:
\begin{align}
&\langle\,ib_0\ \square\chi^0\,\rangle_p = p^3\langle\,\bar{c}_0\ \square C_{12}\,\rangle_p
= -i \frac{p_3^2}{p^2},
\label{eq:Prop-11}\\
&\langle\,i\square b_0\ \chi^0\,\rangle_p = p^3\langle\,\square\bar{c}_0\ C_{12}\,\rangle_p
= -i \frac{p_3^2}{p^2}.
\label{eq:Prop-12}
\end{align}
The nonvanishing inner-products exist between the simple-pole parts and the dipole parts.

Thus we have confirmed that all the fields including multipole fields except for the transverse modes
fall into the BRST quartets and decouple from the physical sector, leaving the two transverse graviton modes
as physical states.

\section{Conclusions}
\label{discussions}

In this paper, combining the new ideas in I and \cite{KNO1}, we have given a new covariant BRST quantization
of UG. First, we have used rank-2 antisymmetric tensor fields for the ghosts which satisfy the transverse
condition. This gives further gauge invariance in the ghost sector and we have successfully gauge fixed TDiff.
For the antighosts, we need fields with the same number of dofs. In I, we have used the same rank-2 antisymmetric
tensor fields, but this leads to a formulation with lots of ghosts and tripole modes in the ghost sector.
Here instead we have used vector antighost which satisfy transverse condition on shell due to the additional
scalar field $S$. We have shown that this leads to a fully gauge-fixed theory, and all the modes except
for the 2 transverse graviton modes fall into the BRST quartets and completely decouple from the physical subspace.
In this way we have succeeded in giving a covariant BRST quantization of UG with the correct number of dofs,
the same as GR.

This formulation has two advantages compared with our previous formulation in I.
One is that the formulation contains less number of ghosts without those originating from antighosts.
The second is that we do not have tripole modes in the ghost sector.
These lead to considerable simplification in the resulting system.
A difference is that this formulation is asymmetric in ghost and antighost sectors,
while that in I is symmetric.
Both formulations give the correct number of dofs.
Which formulation is more useful for covariant study of UG remains to be seen, but we hope that
our formulations should be useful for further study of UG.

\section*{Acknowledgment}

T.K. is supported in part by the JSPS KAKENHI Grant Number JP18K03659.
N.O. is supported in part by the Grant-in-Aid for Scientific Research Fund of the JSPS (C) No. 16K05331,
No. 20K03980, and Taiwan Grant No. MOST 110-2811-M-008-526.


\end{document}